\documentclass[aps, prb,
twocolumn,showpacs]{revtex4}
\usepackage{t1enc}
\usepackage[latin1]{inputenc}
\usepackage[english]{babel}
\usepackage{graphics}
\usepackage{graphicx}
\usepackage{dcolumn}
\usepackage{bm}
\usepackage{color}
\begin{document}

\title{Electronic structure studies of BaFe$_2$As$_2$ by angle-resolved photoemission spectroscopy}
\author{J.\ Fink,$^{1,2}$  S.\ Thirupathaiah,$^1$ R.\ Ovsyannikov,$^1$ H.A.\ D\"urr,$^1$  R.\ Follath,$^1$ Y.\ Huang,$^3$ 
S.\ de Jong,$^3$  M.S.\ Golden,$^3$ Yu-Zhong Zhang,$^4$ H.O.\ Jeschke,$^4$ R.\ Valent\'{\i},$^4$ C.\ Felser,$^5$ S.\ Dastjani Farahani,$^5$  
M.\ Rotter,$^6$ D.\ Johrendt,$^6$ }
\affiliation{
$^1$ Helmholtz-Zentrum Berlin, Albert-Einstein-Strasse 15,12489 Berlin, Germany\\
$^2$ Leibniz-Institute for Solid State and Materials
Research Dresden, P.O.Box 270116, D-01171
Dresden, Germany\\
$^3$Van der Waals-Zeeman Institute, University of Amsterdam, NL-1018XE Amsterdam, The Netherlands\\
$^4$Inst. für Theor. Physik, Goethe-Universität, Max-von-Laue-Straße 1, 60438 Frankfurt, Germany\\
$^5$Inst. für Anorg. Chemie und Anal. Chemie, Johannes Gutenberg-Universität, 55099 Mainz, Germany\\
$^6$Department Chemie und Biochemie, Ludwig-Maximilians-Universität München, 81377 München, Germany}

\date{\today}

\begin{abstract}
We report high resolution angle-resolved photoemission spectroscopy (ARPES) 
studies of the electronic structure of BaFe$_2$As$_2$, 
which is one of the parent compounds of the Fe-pnictide superconductors. 
ARPES measurements have been performed at 20 K and 300 K, 
corresponding to the orthorhombic antiferromagnetic phase and the tetragonal 
paramagnetic phase, respectively. Photon energies between 
30 and 175 eV and polarizations parallel and perpendicular to 
the scattering plane have been used. Measurements of the Fermi surface 
yield two hole pockets at
the $\Gamma$-point and an electron pocket at each of the X-points. 
The topology of the pockets has been concluded from the dispersion of the 
spectral weight as a function of binding energy. 
Changes in the spectral weight at the Fermi level upon variation of the polarization 
of the incident photons yield important information on the orbital character 
of the states near the Fermi level. No differences 
in the electronic structure between 20 and 300 K could be resolved. 
The results are compared with density functional theory band structure calculations for the tetragonal 
paramagnetic phase. 
\end{abstract}
\maketitle

\section{\label{sec:intro} INTRODUCTION}
\begin{figure}
\includegraphics[ angle = -90,width=6 cm]{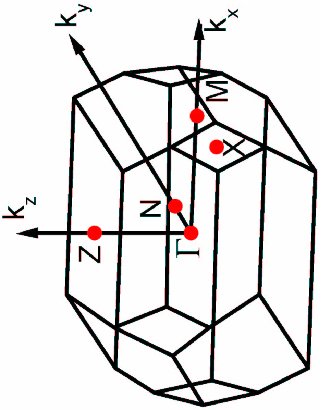}
\caption{(color online) Brillouin zone of BaFe$_2$As$_2$ in the tetragonal phase. The presented ARPES data are focused around
the $\Gamma$ and around the X-point.}   
\label{BZBa122}
\end{figure}

The discovery of high superconducting transition temperatures up to 55 K in iron oxypnictides  
\cite{Kamihara2008, Chen2008,Takahashi2008} has brought a lot of attention to compounds 
containing FeAs layers. Soon thereafter  high superconducting transition temperatures were 
also discovered in the structurally related, non-oxide material
Ba$_{1-x}$K$_x$Fe$_2$As$_2$.\cite{Rotter2008} Similar to the cuprate superconductors, the parent compounds 
of the FeAs-based superconductors have to be doped or, differently from the cuprates,
have to be set under pressure to yield superconductivity. A further difference between 
the cuprates and iron pnictides is that the parent compounds in the latter are not 
antiferromagnetic Mott-Hubbard insulators but metals with an antiferromagnetic ordering. Nevertheless 
superconductivity appears in doped iron pnictides in which antiferromagnetic ordering is
suppressed. There have been many discussions about the relation of a quantum critical point due to magnetic
order and high-$T_c$ superconductivity. In the iron pnictides the antiferromagnetic ordering is supposed to 
occur by a nesting of hole pockets at the center of the Brillouin zone and electron pockets at the zone
corner (X-point in a primitive tetragonal zone).\cite{Mazin2008a} The Brillouin zone for BaFe$_2$As$_2$ is 
presented in Fig.~\ref{BZBa122}.
This nesting scenario may be also important
for the pairing mechanism in these compounds. In order to understand the electronic structure of the new high-$T_c$ 
superconductors it is therefore important to study the electronic structure of the parent compounds.

In this contribution we present the study of the electronic structure of the parent compound of 
the superconductors BaFe$_2$As$_2$ under pressure, Ba$_{1-x}$K$_x$Fe$_2$As$_2$, and BaFe$_{2-x}$Co$_x$As$_2$ using high resolution angle-resolved photoemission spectroscopy (ARPES). Previously other groups have used ARPES to study the electronic structure of these parent compounds and their related superconductors.\cite{Yang2008, Hsieh2008, Yi2009, 
Liu2008, Ding2008, Richard2008, Wray2008, Evtushinsky2008, Liu2008a, Zabolotnyy2008,
Vilmercati2009} 
In the present study we focus on the possible
differences between the electronic structure 
in the paramagnetic tetragonal state and in the antiferromagnetic orthorhombic state of BaFe$_2$As$_2$ by 
performing temperature dependent measurements. This is an interesting topic since the sudden 
decrease of the resistivity below the Neel temperature seen in transport data may indicate that a change occurs in
the electronic structure. A change of the band structure is also predicted from density functional theory (DFT) 
calculations.\cite{Singh2008, Shein2008, Nekrasov2008, Shim2009} Furthermore, we have performed ARPES experiments with 
different polarizations of the photons in order to obtain
information about the orbital character of the bands close to the Fermi level. In addition, using the variable photon energies
available from the synchrotron radiation source, we have obtained information relevant to the $k_z$ values sampled in the measurements, and also are able to make an initial examination of the dispersion of the bands perpendicular to the
FeAs planes. The experimental results are compared with our DFT band structure calculations and with similar work in the literature on doped and undoped
BaFe$_2$As$_2$.\cite{Singh2008, Liu2008, Shein2008, Nekrasov2008, Ma2008} 

\section{\label{sec:exper}  EXPERIMENTAL DETAILS}
Single crystals of BaFe$_2$As$_2$ were grown out of a Sn flux in Amsterdam and M\"unchen, 
using conventional high temperature solution 
growth techniques. Elemental analysis of the former was performed using wavelength 
dispersive X-ray spectroscopy (WDS).
Further elemental analysis was obtained from X-ray induced photoemission spectroscopy on the core levels.\cite{deJong2009} 
Both methods yielded a Sn contamination of the crystal of approximately 1.6 atomic \%. 
According to a recent study of 
BaFe$_2$As$_2$ single crystals, such a Sn contamination leads to a reduction of the structural 
transition temperature from the tetragonal to the orthorhombic phase and also
to a reduction of the transition between the paramagnetic and the antiferromagnetic phase.\cite{Ni2008}
Polycrystalline Sn-free samples show transition temperatures 
close to 140 K. For the Amsterdam crystals a transition temperature of  $\approx$ 65 K 
was deduced from resistivity measurements. Therefore they are termed BFA65K.
For the M\"unchen crystals a structural 
transition and a Neel temperature
at $\approx$ 100 K are derived from neutron scattering data \cite{Su2008} and therefore they are termed BFA100K. 
The higher transition temperature of this crystal when compared to the 
BFA65K crystals indicates a lower Sn content. 
 
\begin{figure}
\includegraphics[width=5 cm]{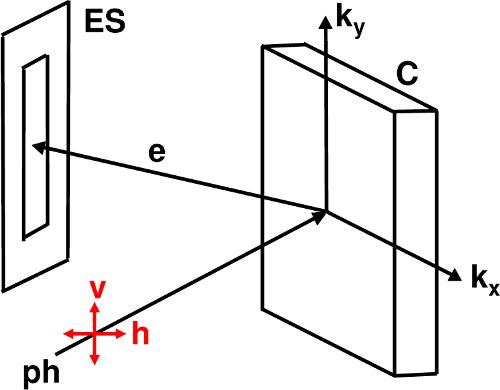}
\caption{(color online) measuring geometry for the ARPES data taken near the $\Gamma$-point. The entrance 
slit (ES) of the analyser, 
the crystal (C), the vertical (v) and the horizontal (h) linear polarization of the incoming photons (ph), 
and the trajectory of the outgoing photoelectrons (e)
are shown. The direction of the photons and the photoelectrons define the scattering plane. This scattering
plane is almost horizontal for photoelectrons traveling to the center of the entrance slit (k$_y$=0). }   
\label{Geometry}
\end{figure}

The ARPES experiments were carried out at the BESSY synchrotron
radiation facility using the U125/1-PGM beam line and the ''1$^3$-ARPES'' end station provided with a SCIENTA
R4000 analyzer. Spectra were taken with various photon energies
ranging from h$\nu$ = 30 to 175 eV. The total energy resolution ranged from
10 meV (FWHM) at photon energies h$\nu$ = 30 eV to 20 meV at
h$\nu$ = 175 eV.  The angular resolution was 0.2° along the slit of the analyzer and 0.3° perpendicular to it.
The experimental geometry is depicted in Fig.~\ref{Geometry} and is also described in more detail 
in a previous paper. \cite{Inosov2008}
In this geometry the scattering plane, as defined by the direction of the incoming photons and the trajectory of the outgoing
photoelectrons, is nearly horizontal for k$_y$ = 0, i.e., for photoelectrons traveling to the
center of the entrance slit of the analyzer. 

The present ARPES study was mainly focused to two 
regions in the Brillouin zone: round the zone center (the $\Gamma$ = (0,0,0) and Z = $\frac{\pi}{c}$(0,0,1) points) 
and the zone corner (X = $\frac{\pi}{a}$(1,1,0) point), as illustrated in Fig.~\ref{BZBa122}, where $a$ and $c$ are the 
tetragonal lattice constants for BaFe$_2$As$_2$ along x and the z-axis, respectively.
For recording data at the $\Gamma$-point we oriented the crystal in such a way that the $k_x (k_y)$ direction was parallel to
the $\Gamma$-M(N) direction, where M = $\frac{\pi}{a}(1,0,0)$, N = $\frac{\pi}{a}(0,1,0)$. In this way the $\Gamma$MZ-mirror 
plane of the crystal is in the scattering
plane. Using a helical undulator,
the linear polarization of the incoming synchrotron radiation could be changed from the horizontal 
direction to the vertical direction.
Due to symmetry selection rules within the matrix elements governing photoemission, the experimental 
intensities are strongly affected by the 
orientation of the polarization of the photons relative to the scattering plane.\cite{Huefner1994}
For $k_y = 0$, i.e., for photoelectrons traveling to the center of the exit slit, vertical (horizontal) polarization 
means that the electric field vector is essentially perpendicular (parallel) to 
the scattering plane, 
yielding the label s(p)
for the two polarization geometries. We emphasize that due to the finite 
size of the vertically aligned entrance slit this only holds for the center of this slit.
Orienting a mirror plane of the crystal into the scattering plane and performing polarization 
dependent ARPES experiments, important information on the parity of orbitals relative to the mirror pane can be
obtained. Since the final state is even with respect to reflection in the mirror plane, the matrix element 
and thus the intensity should vanish when
the product of the dipole operator and the initial state is odd. This means that
the intensity from initial state orbitals having odd(even) parity relative to the mirror plane should disappear for
parallel(perpendicular) photon polarization (or p[s] polarizations) relative to the mirror plane.
 
For recording data at the X-point the crystal was rotated around its normal by 45°. The horizontal (vertical) direction, which 
we now term $k_{x'}$($k_{y'}$), is parallel to the $\Gamma$-X(Y) direction where Y = $\frac{\pi}{a}(-1,1,0)$. 
In this way the  $\Gamma$XZ-mirror
plane lies in the scattering plane for $k_{y'}$ = 0.

Finally, for exploring the photon energy dependence of the data and for elucidating the $k_z$ values 
sampled under our experimental 
conditions ($k_z$ values near the $\Gamma$-point, parallel to $\Gamma$-Z, we recorded the 
kinetic energy of the photoelectrons in nearly normal emission for various $k_y$ values (along the $\Gamma$-N direction) 
as a function of the photon energy.

Samples were mounted on a high-precision cryomanipulator
and cleaved in situ at room temperature in an ultrahigh-vacuum chamber with a
base pressure of 10$^{-10}$ mbar.

Most of the data presented here were taken from the larger BFA65K crystals. In some cases (marked in the 
figure captions), we also present data from the much smaller BFA100K crystals.

\section{\label{sec:band}  band structure CALCULATIONS}
For comparison with ARPES data we calculated the electronic
band structure of tetragonal paramagnetic BaFe$_2$As$_2$
within the DFT in the generalized
gradient approximation (GGA)~\cite{Perdew1996} using the full-potential
linearized augmented plane-wave code WIEN2k.~\cite{Blaha2001a} We considered
a $k$ mesh of $40\times 40\times 40$
in the irreducible Brillouin zone and an accuracy benchmark for the LAPW (Linear augmented 
plane wave basis) of Rk$_{max}$ = 7. 
We performed calculations (i) by fixing the As position to the experimental
value~\cite{Huang2008} (ii) by fully relaxing the lattice parameters as well as the internal As coordinate within
spin-polarized GGA.~\cite{Zhang2008} 
Good agreement with ARPES is only obtained when the As position is kept as given by 
the neutron diffraction data,~\cite{Huang2008} as presented in this work or by performing GGA calculations on the optimized
structure with tetragonalization.~\cite{Zhang2008}

\begin{figure}
\includegraphics[ width=7 cm]{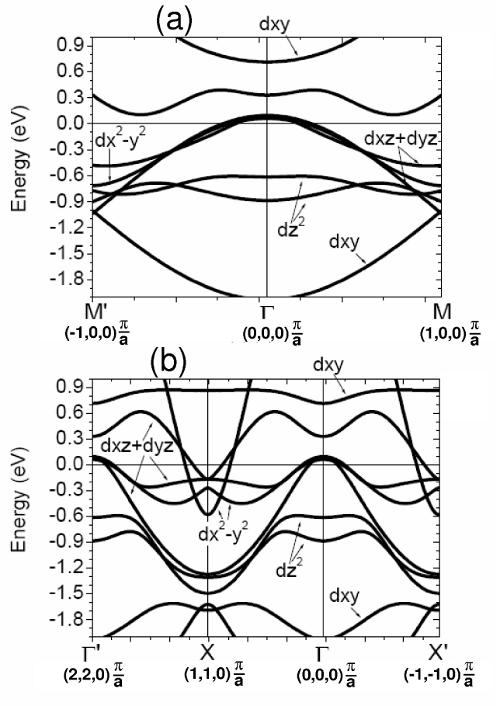}
\caption{band structure calculation for tetragonal paramagnetic BaFe$_2$As$_2$ for the energy region close to the Fermi level 
including an assignment of the dominant Fe $3d$ orbital character of the bands in the coordinate frame
$x = a$ and $y = b$.
(a) The $\Gamma$-M direction. (b) The $\Gamma$-X direction.}   
\label{Band}
\end{figure}

\begin{figure}
\includegraphics[angle = -90,width=7 cm]{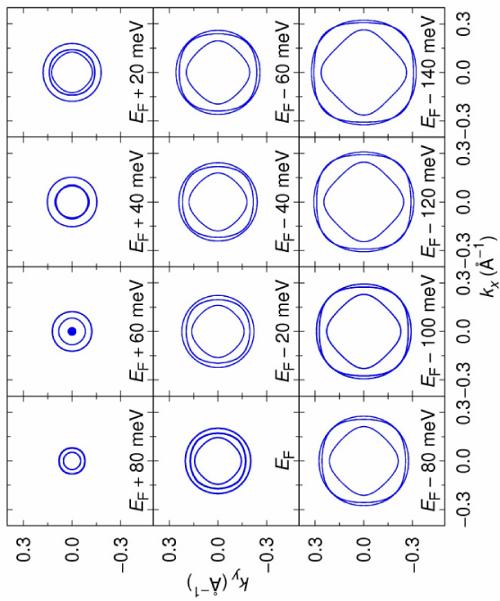}
\caption{(color online) calculated constant energy contours for tetragonal paramagnetic BaFe$_2$As$_2$ around the
 $\Gamma$-point for k$_z$=0. k$_x$ is along the $\Gamma$-M direction.}  
\label{Gamma}
\end{figure}

\begin{figure}
\includegraphics[ angle=-90,width=7 cm]{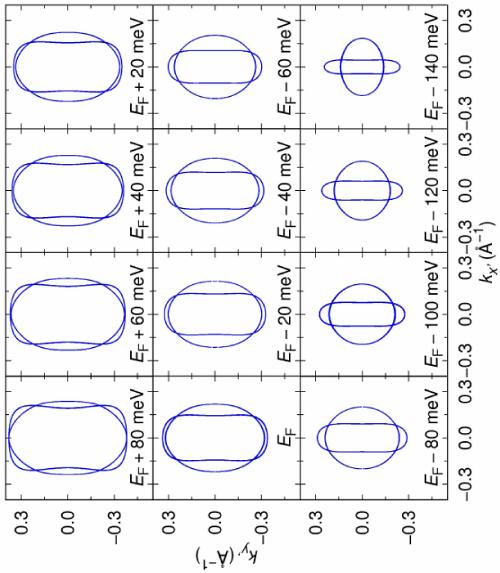}
\caption{(color online) calculated constant energy contours for tetragonal paramagnetic BaFe$_2$As$_2$ around the
X-point for k$_z$=0. k$_{x'}$ is along the $\Gamma$-X direction.}    
\label{Ex}
\end{figure}

In Fig.~\ref{Band} we show the calculated band structure for BaFe$_2$As$_2$ close to the
Fermi level along the
$\Gamma$-M and the $\Gamma$-X direction.
In Fig.~\ref{Gamma} and  Fig.~\ref{Ex} we depict calculated constant energy contours for various binding energies
close to the $\Gamma$- and the X-point, respectively. 

\section{\label{sec:res}  RESULTS}

\begin{figure}
\includegraphics[width=7 cm]{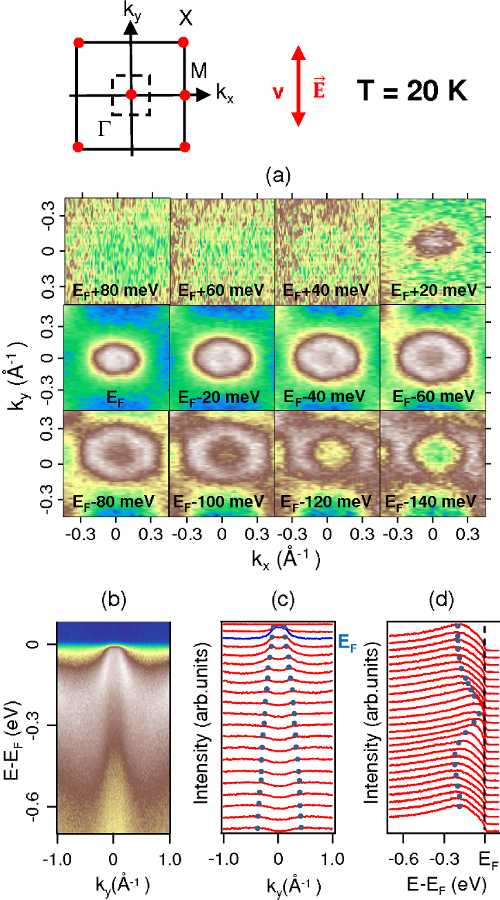}
\caption{(color online) ARPES data around the center of the Brillouin zone ($\Gamma$-point)
of BaFe$_2$As$_2$ (BFA65K crystal) measured with a photon energy h$\nu$ = 75 eV. The top line shows the two-dimensional
Brillouin zone together with the scanning range (dashed line). It also gives the polarization of the photons
(in this case vertical and parallel to $k_y$) and the recording temperature (here 20 K).  
(a) Momentum distribution maps for different binding energies derived by an integration over an energy range of
10 meV. (b) Spectral intensities as a 
function of $k_{y}$ and binding energy for $k_{x}$ = 0.
(c) Momentum distribution curves for for $E - E_F$ = 50 to -700 meV.
(d) Energy distribution curves for wave vectors parallel to $k_{y}$ ranging from
-1 {\AA}$^{-1}$ to 1 {\AA}$^{-1}$.}    
\label{20KvertG}
\end{figure}

\begin{figure}
\includegraphics[width=7 cm]{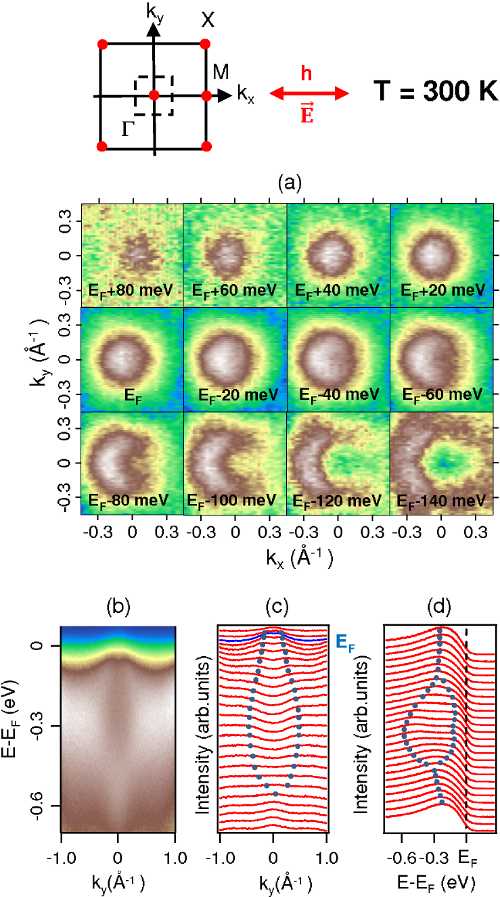}
\caption{(color online) analogous data as in Fig.~\ref{20KvertG} but now recorded with a horizontal photon polarization 
(parallel to $k_x$) and at a temperature $T$ = 300K.}    
\label{300KhoriG}
\end{figure}

\begin{figure}
\includegraphics[width=5 cm]{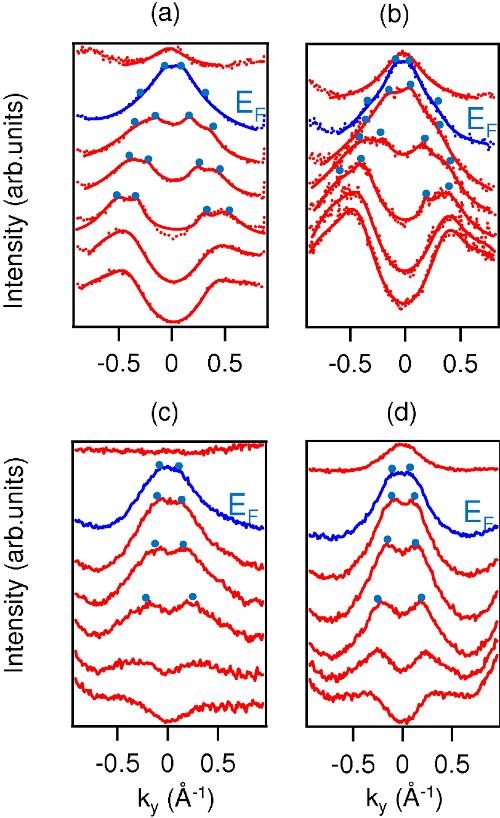}
\caption{(color online) ARPES momentum distribution curves for $E-E_F$ = 30 to 150 meV near 
the $\Gamma$-point for $k_x$ = 0 along  the $k_y$
direction for a BFA100K and a  BFA65K crystal using horizontally polarized photons with energies 
h$\nu$ = 50 eV and h$\nu$ = 75 eV, 
respectively. The solid lines are based on a fit to the data.
A geometry was used as depicted in Fig.~\ref{300KhoriG}.
(a) BFA100K crystal at $T$ = 20 K. (b) BFA100K crystal at $T$ = 300 K.
(c) BFA65K crystal at $T$ = 20 K. (d) BFA65K crystal at $T$ = 300 K.}    
\label{20KhoriGMun}
\end{figure}

In Fig.~\ref{20KvertG} we show ARPES data of BaFe$_2$As$_2$ near the $\Gamma$-point recorded at a temperature of
20 K using photons polarized parallel to the vertical direction, i.e., s-polarization for a horizontal scattering plane. 
In Fig.~\ref{20KvertG} (a) we depict momentum distribution maps for various binding energies. 
We remark here that we also present data for energies above the Fermi level since due to thermal excitation, states
above $E_F$ can be populated and thus can be detected by ARPES. In the momentum distribution map for $E=E_F$ 
an almost circular Fermi surface is realized. 
The increasing diameter with increasing binding energy indicates clearly that this Fermi surface is 
caused by a hole pocket centered at the $\Gamma$-point. A Fermi wave vector 
$k_F^y = 0.06 \pm 0.01 $ {\AA}$^{-1}$
and a  Fermi velocity $v_F^y = 0.85 \pm 0.04 $ eV{\AA} along the $k_{y}$ direction is estimated from the data shown in
Fig.~\ref{20KvertG}. Along the x-direction the Fermi wave vector is slightly larger 
($k_F^x = 0.11 \pm 0.01 $ {\AA}$^{-1}$).
In Fig.~\ref{300KhoriG} we show analogous data as in Fig.~\ref{20KvertG}, but now 
taken with photons polarized along the horizontal
direction, i.e., p-polarized radiation for a horizontal scattering plane (k$_y$ = 0). The measurement temperature was
T = 300 K, well into the paramagnetic, tetragonal phase. Here, a circular Fermi surface is observed. The increasing
size of the constant energy contours with increasing binding energy again indicates that at 
the $\Gamma$-point we have hole pockets.
The Fermi wave vector and the Fermi velocity  is
$k_F = 0.10 \pm 0.01$ {\AA}$^{-1}$
and $v_F = 0.72 \pm 0.05 $ eV{\AA}, respectively. Contrary to the data taken
with vertical photon polarization  at 
$E-E_F \approx$ -600 meV spectral weight appears in the center of the zone. 
This indicates that there is now a band appearing which 
has a non-zero matrix element. 
Almost identical data were obtained at $T$ = 20 K (not shown), which gave $k_F = 0.11 \pm 0.01$ {\AA}$^{-1}$
and $v_F = 0.72 \pm 0.05 $ eV{\AA}. A remarkable result is that 
the Fermi surface, the Fermi wave vector, and the Fermi 
velocity do not change between $T$ = 20 K and $T$ = 300 K within error bars.
In Fig.~\ref{20KhoriGMun} we compare momentum distribution curves close to the Fermi level from a BFA100K crystal and from a
BFA65K crystal recorded at $T$ = 300 K and $T$ = 20 K. 
In the BFA100K curve taken at low temperatures, 
clearly two hole pockets can be distinguished. This agrees with previous ARPES studies.\cite{Liu2008, Yang2008}
The corresponding Fermi wave vectors of the two bands are 
$k_F = 0.10 \pm 0.01 $ {\AA}$^{-1}$ and $k_F = 0.27 \pm 0.01 $ {\AA}$^{-1}$
and  the corresponding  Fermi velocities $v_F = 0.69 \pm 0.06 $ eV{\AA} and $v_F = 0.75 \pm 0.05 $ eV{\AA} are estimated along the $k_{y}$ direction. At high temperatures it is more difficult to resolve the two bands  
due to the increased thermal broadening. Nevertheless $k_F$ values for the two Fermi surfaces could be derived: for the inner 
and the outer 
Fermi surface we obtain $k_F = 0.09 \pm 0.01 $ {\AA}$^{-1}$ and $k_F = 0.28 \pm 0.01 $ {\AA}$^{-1}$, respectively. 
The corresponding Fermi velocities are $v_F = 0.72 \pm 0.08 $ eV{\AA} and $v_F = 0.73 \pm 0.06 $ eV{\AA},
respectively. The curves from the BFA65K crystal are similar but slightly broadened possibly 
due to the higher Sn content which
may lead to an additional scattering related to the impurity scattering. 
Therefore it is difficult to resolve contributions from the
outer Fermi surface and the Fermi wave vectors and the Fermi velocities for the BFA65K sample given above 
are mainly related to the inner Fermi surface.

\begin{figure}
\includegraphics[width=7 cm]{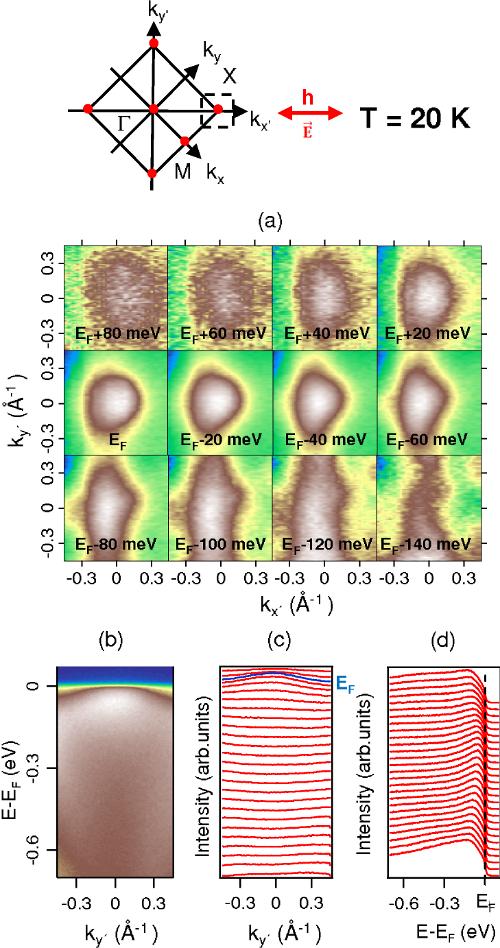}
\caption{(color online) similar data as in Fig.~7 measured with a photon energy h$\nu$ = 75 eV 
but now around the X-point and at a temperature was $T$ = 20 K. Horizontal photon polarization (parallel to $k_{x'}$) 
has been used. The horizontal $k_{x'}$ direction is aligned parallel to the $\Gamma$-X direction.}    
\label{20KhoriM}
\end{figure}

\begin{figure}
\includegraphics[width=7 cm]{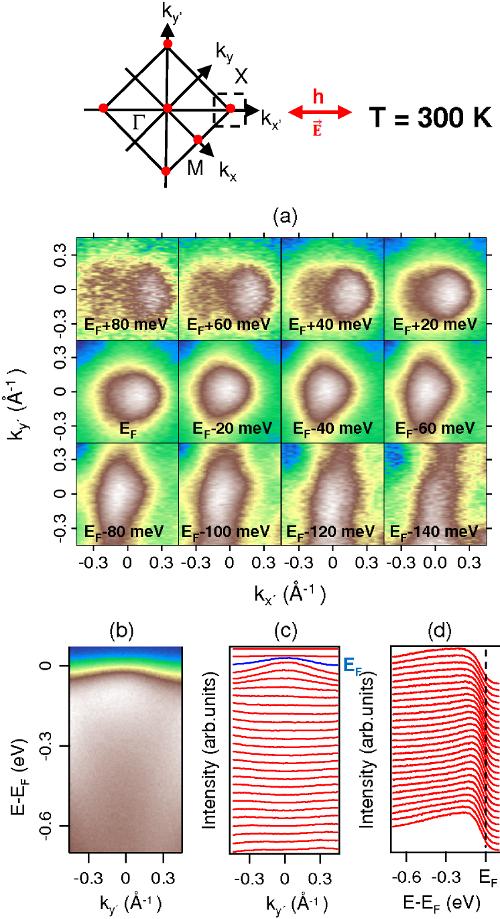}
\caption{(color online) similar data as in Fig.~\ref{20KhoriM} around the X-point but now with the recording
temperature $T$ = 300 K. Again a horizontal photon polarization 
has been used.}    
\label{300KhoriM}
\end{figure}

\begin{figure}
\includegraphics[width=7 cm]{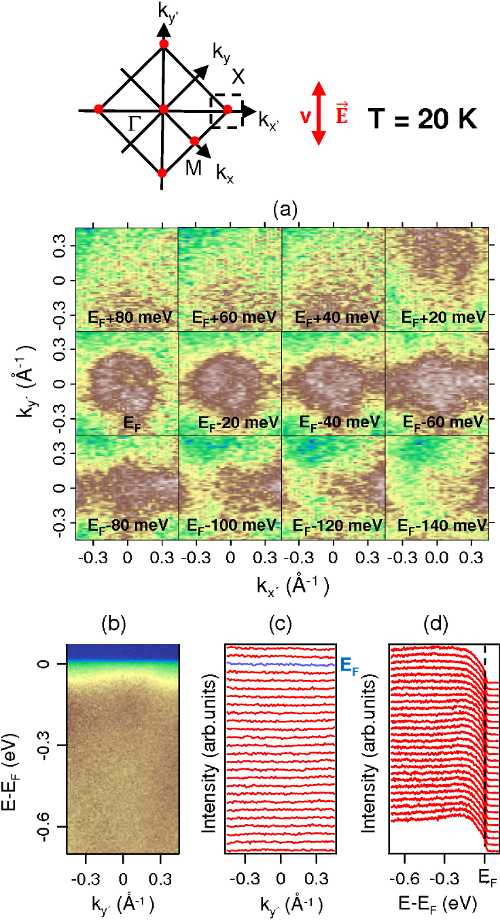}
\caption{(color online) similar data as in Fig.~\ref{20KhoriM} taken at a temperature of 20 K but with vertical 
photon polarization (parallel to $k_{y'}$).}    
\label{20KvertM}
\end{figure}

In the following we present data taken near the X-point. Now the horizontal $k_{x'}$ direction is parallel to 
$\Gamma$-X, i.e., the crystal has been rotated by 45° around the surface normal (see Section~\ref{sec:exper}). In 
Fig.~\ref{20KhoriM} we show data 
taken at$T$ = 20 K 
with horizontal photon polarization. This means that in the center of the analyzer slit ($k_{y'}$ = 0) we have p-polarization. 
While close to the Fermi level there is more a circular spectral weight distribution,
at lower energy ($E-E_F$ < -60 meV) an elongated propeller-blade like distribution along the $k_{y'}$ direction is realized.
In Fig.~\ref{300KhoriM} we present analogous data but taken at $T$ = 300 K. 
Essentially the data are very close to the low temperature data.
In similar spectra but now taken with vertical polarization at $T$ = 20 K, shown in Fig. 11, a circular 
Fermi surface with a Fermi wave 
vector $k_F = 0.16 \pm 0.02 $ {\AA}$^{-1}$ is detected while at higher binding energies ($E-E_F$ < -60 meV) 
an elongated blade-like 
distribution along the $k_{x'}$ direction is visible (see Fig.~\ref{20KvertM}). As will be outlined in 
Section~\ref{sec:dis} shortly, the 
differences between data recorded using the two polarizations can be attributed to matrix element effects 
(selection rules), which enable us to conclude about the orbital character of the states involved. 

\begin{figure}
\includegraphics[width=7 cm]{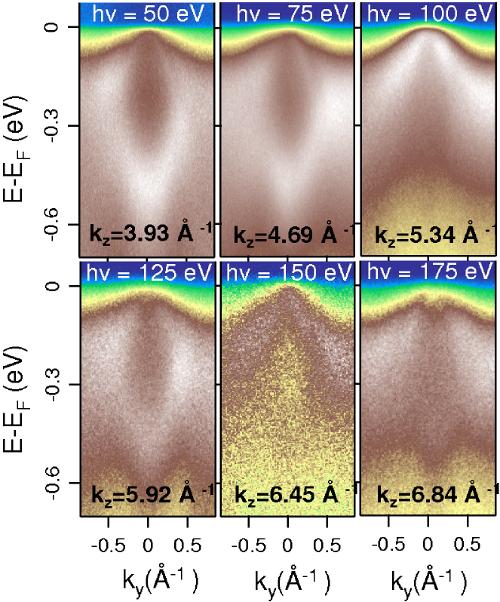}
\caption{(color online). ARPES data from a BFA100K crystal recorded around the center of the Brillouin zone ($\Gamma$-point)
measured with horizontal photon polarization and variable photon energies 
ranging from h$\nu$ = 50 to 175 eV. Data are recorded in normal emission at a temperature of $T$ = 20 K. $k_x$ is parallel
to the $\Gamma$-M direction.}    
\label{kz}
\end{figure}

Finally, in Fig.~\ref{kz} we present cuts along the $\Gamma$-N  direction (parallel to k$_y$) for 
various photon energies and normal photoelectron emission recorded from a BFA100K crystal at T= 20 K. 
This data is very useful in the context of the comparison between the ARPES and DFT data, as \emph{via} the 
identification of periodicities (vs. photon energy) we are able to cross-check our position in $k_z$, as in 
this measurement, the wave vector parallel to the
FeAs layers is directed along the $k_{y}$ direction and the $k_z$ values are varied as the photon energy changes. 
The most clear result visible in Fig. 12 is that, near $E-E_F$ = -600 meV, a remarkable reduction of 
spectral weight is observed for the
photon energies h$\nu$ = 100 and 150 eV. This effect will be discussed in Section~\ref{sec:dis}.

\section{\label{sec:dis}  DISCUSSION}

First we discuss the results obtained near the $\Gamma$-point shown in Figs.~\ref{20KvertG} to \ref{20KhoriGMun}.
According to the present band structure
calculations and those published in the literature \cite{Singh2008, Shein2008, Nekrasov2008,Shim2009} for the 
paramagnetic tetragonal state the in-plane Fermi
surface at the $\Gamma$-point should be caused by three hole pockets formed by three bands which are 
almost degenerate (see e.g. Fig.~\ref{Band}(a) and the panel $E_F$ in Fig.~\ref{Gamma}). The orbital character of 
the three bands at the $\Gamma$ and at the Z-point is predicted to be related to mainly  
Fe $3d_{x^2-y^2}$, Fe $3d_{yz}$, and Fe $3d_{xz}$ states.
In the calculations, these bands show
dispersion along the $\Gamma$-Z direction which leads to a departure from a simple cylindrical 
form of the Fermi surface as a function of $k_z$ value. For $k_z$ = 0 the Fermi wave vectors 
for the Fe $3d_{x^2-y^2}$, Fe $3d_{xz}$, and 
Fe $3d_{yz}$ bands are 0.17, 0.14, and 0.20 {\AA}$^{-1}$, respectively.  
For a $k_z$ value at the zone boundary, the splitting between the three bands is increased
and essentially two larger Fermi surfaces with  $k_F$ values of 0.22 and 0.42 {\AA}$^{-1}$
are predicted.  At the $\Gamma$-point
two further bands with mainly
Fe $3d_{z^2}$ character appear near $E-E_F$ = -600 and -900 meV.
For $k_z$-values between the $\Gamma$ and the Z-point these bands disperse toward  the Fermi level and reach an 
energy of $\approx $ -100 meV.

As described in Section II, due to matrix element effects for horizontal
(vertical) polarization of the incoming photons and a horizontal scattering plane, odd(even) 
initial states relative to the $\Gamma$MZ mirror plane should disappear. For the data shown in 
Fig.~\ref{20KvertG}, where the photons were polarized vertically (s-polarization), all even states relative to the
mirror plane, i.e., Fe $3d_{x^2-y^2}$, Fe $3d_{z^2}$, and  Fe $3d_{xz}$, should disappear along the horizontal $k_y$ = 0 line, 
but the other two odd Fe $3d$ orbitals (Fe $3d_{xy}$ and Fe $3d_{yz}$) should
contribute to the spectral weight along this line. Since the Fe $3d_{xy}$ bands are outside the
energy range presented in these measurements only the Fe $3d_{yz}$ band contributes along the horizontal $k_y$ = 0 line.
For data taken with horizontal photon polarization (p-polarization, presented in Fig.~\ref{300KhoriG}), and for a cut through the data with $k_y$=0, the odd Fe $3d_{yz}$ (and $3d_{xy}$) states should be 
suppressed, but all other Fe $3d$ bands near the Fermi should be visible.

As pointed out before in all data taken at the center of the two-dimensional Brillouin zone 
(see Figs.~\ref{20KvertG} to \ref{20KhoriGMun}), one clearly 
sees one or two hole pockets, i.e., Fermi surfaces in the 
$k_{x},k_{y}$ plane, the diameter of which decreases when going from below to above the Fermi level.
It is remarkable that the diamond-like constant energy contour predicted in Fig.~\ref{Gamma} for the 
Fe $3d_{x^2-y^2}$ states at higher binding energies is well reproduced in the experimental data.

In the following we attempt a more quantitative comparison of the ARPES data near the Fermi energy with 
the band structure calculations. 
We warn the reader in advance that a fully self-consistent assignment that matches the data 
one-on-one with the LDA bands is not achievable: most probably a result of the multiband nature of 
these materials, the uncertainty experimentally about the exact value of $k_z$ and the fact that knowledge about 
the k, E and polarization dependence of the photoionisation matrix elements in these systems is only just 
starting to be gathered.

Since in the data shown in Fig.~\ref{20KvertG} 
near the $k_y$ = 0 line we should only
see Fe $3d_{yz}$ states near the Fermi surface, the experimental value $k_F^x = 0.11 \pm 0.01 $ {\AA}$^{-1}$ 
should be compared with
the theoretical value $k_F$ = 0.20 {\AA}$^{-1}$ for the Fe $3d_{yz}$-derived hole pocket. For the  $k_y$ direction 
we should observe the 
same value as along the $k_x$ direction, 
but in the experiment a considerably smaller value $k_F^y = 0.06 \pm 0.01 $ {\AA}$^{-1}$ is observed. An explanation for
this reduced Fermi wave vector along the $k_y$ direction could be that moving to non-zero $k_y$ values, 
also spectral weight should appear from 
one or both of the two low energy even-symmetry bands formed by overlap between the Fe $3d_{xz}$ and Fe $3d_{x^2-y^2}$ states.
Assuming smaller $k_F$ values for one or both of these even symmetry bands could result in an apparent asymmetric Fermi surface
with a smaller diameter along the $k_y$ direction. With a view to the band structure data for $k_z$=0, 
we tentatively assign the Fermi surface with the small $k_F^y$ value to the Fe $3d_{x^2-y^2}$-related states.

In the data recorded with horizontally polarized light shown in Figs.~\ref{300KhoriG} and  \ref{20KhoriGMun}, 
the outer Fermi surface with 
$k_F = 0.27 \pm 0.01 $ {\AA}$^{-1}$ can be tentatively assigned to the Fe $3d_{xz}$ states. 
In this way we 
can arrive at  qualitative but not
quantitative agreement with the band structure calculations, assuming our ARPES data with 75eV photons to be 
representative for $k_z$=0 (we will present arguments as to why we believe this to be the case later on). 
The experimental $k_F$ values are 
somewhat different from the calculated ones
and the experimentaly observed splitting of the three bands is about a factor three larger than the calculated one. 
For the spectra taken 
with horizontal photon polarization (see Fig.~\ref{300KhoriG}), the experimental Fermi surface near the
$k_y$ = 0 line should be caused by the even Fe $3d_{x^2-y^2}$ and Fe $3d_{xz}$ bands. Thus, based on the 
assignment made above, a small Fermi wave vector of ca. $k_F = 0.06  $ {\AA}$^{-1}$ is expected here 
too (the Fe $3d_{x^2-y^2}$ band), yet in these data we find experimentally a value of $k_F = 0.11 \pm 0.01$ {\AA}$^{-1}$. 
The observation of an even larger second Fermi surface in the data shown
in Fig.~\ref{20KhoriGMun} with a Fermi vector $k_F = 0.27 \pm 0.01 $ {\AA}$^{-1}$ which consequently would 
need to be assigned to the 
Fe $3d_{xz}$ bands, a fact which would not match with the interpretation we proposed for the  
asymmetric Fermi surface detected for vertical photon polarization shown in Fig.~6.
This discussion serves to illustrate that further work is necessary to obtain an unambiguous assignment of 
the detected bands to the results of DFT band structure calculations.

Taking the lack of one-to-one quantitative agreement between the ARPES data and the band structure calculations at face value, we now mention and comment on the following possible origins for these differences - at least for the states close to $E_F$.
(i) The surface doping level and/or electronic structure is different to that for the bulk. This could be due to the formation of polar surfaces connected with a small shift of the chemical potential, or due to reconstructions following on from the significant disorder seen in the Ba termination layer of these crystals using STM,\cite{Yin2008, 
Massee2008,Hsieh2008} which could then change the electronic structure of the FeAs block closest to the surface. A further source of a surface/bulk difference could be a relaxation of the As-Fe-As distances at the surface, as we know from our DFT calculations that the electronic structure of these compounds is very sensitive to the Fe-As distance.
However, a scenario involving a different surface electronic structure compared to the bulk is constrained by a recent high-energy, angle-integrated photoemission spectroscopic study conducted by some of us on BaFe$_2$As$_2$, which concludes that there are, at most, minor alterations of the electronic structure at the surface compared to the bulk.\cite{deJong2009}
(ii) A second scenario would involve the existence of a large in-plane antiferromagnetic correlation length in the 
paramagnetic state of BaFe$_2$As$_2$ which may shift the electronic states with respect to the purely paramagnetic state assumed in the calculations, and would also provide a basis for the essential lack of temperature dependence seen in the experimental data.
(iii) Thirdly, from the experimental point of view, we have an uncertainty in the $k_z$ values. This scenario suffers from the drawback that the diameter of our largest $\Gamma$-centered Fermi surface is smaller (by a factor of almost one half) than the maximal DFT-predicted value at the Z-point: thus even allowing ourselves a totally free choice of our true $k_z$-value in experiment, there still remain discrepancies between the ARPES and DFT data. In addition to this point, we present arguments below in favor of the fact that we know the $k_z$-location of the data recorded with the photon energies 50 and 75 eV to be close to zero (i.e. at the $\Gamma$-point).

Next we discuss the Fermi velocities derived for the Fermi surfaces near the $\Gamma$-point. From the band 
structure calculations
we obtain values between 0.7 and 1.1 eV{\AA} for the three bands along the $\Gamma$-M direction. 
This is in fair agreement with the
experimental values $v_F \approx $ 0.8 eV{\AA} along the same direction determined from spectra taken 
for vertical and horizontal
photon polarization. Thus the present measurements do not indicate a large mass renormalization of the bands 
near the $\Gamma$-point. On the other hand, with the present energy resolution we cannot exclude a mass renormalization 
due to bosonic excitations
with an energy less than $\approx$ 20 meV, since the bands should be renormalized only below the energy of the bosonic excitations.

At higher binding energies we obtain nice agreement between the band structure calculations 
and the experimental data presented in 
Figs.~\ref{20KvertG} and \ref{300KhoriG}. The Fe $3d_{z^2}$ band, which according to Fig.~\ref{Band} should 
appear near an $E-E_F$ value of -600 meV, is clearly visible in the data taken with horizontal polarization 
shown in Fig.~\ref{300KhoriG}(b), which is consistent with the orbital origin of these states which is always even with respect 
to the $\Gamma$MZ mirror plane. On the other hand for vertical photon polarization, these states should disappear, 
which is indeed exactly what they do, as can be seen from the data shown in  Fig.~\ref{20KvertG}(b).
The fact that such an unambiguous assignment of these 600meV states for the p-polarisation geometry is 
possible, presents us with a clear and thus powerful 'tool' to gauge the position of our ARPES data 
with respect to the relevant value of $k_z$. As mentioned earlier, for $k_z$=0, these states should lie at ca. 600meV. 
They are predicted (within DFT) to disperse up to only within 100 meV of $E_F$ for $k_z$ values at the bottom or 
top of the Brillouin zone. This means, therefore, that the observed binding energy (600meV) of these Fe $3d_{z^2}$ 
states for $k_x$=$k_y$=0 thus 'pins' the data recorded using 75eV photons to $k_z$ values at or close to zero.

At the end of the discussion of the data at $\Gamma$(Z) we point out that we did not observe a significant
change of the electronic structure when going from the tetragonal paramagnetic phase at high temperatures to
the orthorhombic antiferromagnetic phase at low temperatures. This observation is in line with previous ARPES
studies\cite{Yang2008} but not with previous spin polarized DFT band structure calculations\cite{Ma2008,Zhang2008} 
which predicted a large 
change of the electronic structure near the $\Gamma$-point in the orthorhombic antiferromagnetic phase. 
On the one hand, this difference could possibly be attributed to the strong itineracy of the BaFe$_2$As$_2$ system which causes only small changes of the electronic structure, while spin polarized DFT calculations overestimate the magnetic moment and thus 
predicting large changes of the electronic structure. On the other hand, as alluded to earlier, there also exists the possibility that even above $T_N$ strong quasi-2D antiferromagnetic spin-spin correlations (of reasonably long range) are to be found  within the FeAs layers. This would mean that the two-dimensional electronic structure within these layers is only very weakly altered  when crossing the bulk, three-dimensional spin-ordering temperature, $T_N$.

In the following we discuss the results around the X-point. According to the 
band structure calculations (see Figs.~\ref{Band} and \ref{Ex}) the in-plane Fermi
surfaces around this high-symmetry point should be caused by two electron pockets. 
The larger one is mainly formed by Fe $3d_{x^2-y^2}$ states while the smaller one has mainly Fe $3d_{xz}$ and Fe 
$3d_{yz}$ character. Measuring from X  
towards the $\Gamma$-point, the
$k_F$ values are predicted to be 0.14 and 0.22 {\AA}$^{-1}$. The band-bottoms of the small and the large electron 
pockets are predicted by DFT to appear at $E-E_F$ = 
-180 meV and -550 meV, respectively. At -180 meV, a third weakly dispersing band appears in
the calculations which has mainly Fe $3d_{xz}$ 
and Fe $3d_{yz}$ character. A further band with a larger dispersion having mainly Fe $3d_{x^2-y^2}$ character is predicted 
at around -300 meV. 
 
In all data taken near the X-point, the crystal was oriented in such a way that a $\Gamma$XZ mirror plane  
is in the scattering plane for the center of the exit slit of 
the analyser ($k_{y'}=0$). In the data for vertical (i.e. s) polarization shown in Fig.~\ref{20KvertM},
an almost spherical Fermi surface is detected around the X-point, with a Fermi 
vector $k_F$ = 0.16 $\pm $ 0.02 {\AA}$^{-1}$. For this photon polarization, the experimentally observed
Fermi surface could be related to all three bands close to the Fermi level having Fe $3d_{yz}$, 
Fe $3d_{xz}$, and Fe $3d_{x^2-y^2}$ character. The experimental $k_F$ value is not far from that 
predicted from the bandstrucure calculations for the smaller (Fe $3d_{yz}$, Fe $3d_{xz}$) electron pocket.

According to the calculations presented in Fig.~\ref{Ex}, with increasing binding 
energy the constant energy surfaces around the X-point should change shape quite significantly. 
The elliptical energy surface (with its long axis pointing along $k_{y'}$) increases in aspect ratio significantly 
as the binding energy increases, until these states finally disappear near -180 meV. The second Fermi 
surface seen in Fig. 5 is due to the Fe $3d_{x^2-y^2}$ states, and transforms with increasing binding 
energy from an ellipse with its long axis pointing along $k_{y'}$ at $E_F$ to a more rounded form, 
slightly elongated along $k_{x'}$ at higher binding energies. According to the DFT data, the band-bottom 
for these states should lie at an $E-E_F$ of ca. -600 meV, with another (quite flat) band of the same symmetry 
bottoming out at roughly half this value.

The experimental results shown in Fig.~\ref{20KvertM}(a), although hampered by the low 
cross-section for these conditions, show a small, rounded Fermi surface, transforming into a blade-like 
intensity distribution with its long axis along $k_{x'}$, before the intensity essentially runs out at -140 meV. 
Given the s-polarisation used to record this data, we would expect to be sensitive to the Fe$3d_{x^2-y^2}$ 
states along the $k_{y'}=0$ line, thus the fact that the constant energy cuts broaden parallel to $k_{x'}$ with 
increasing binding energy is an aspect that matches with the DFT predictions. It is less clear at present 
why the observed intensity for $E=E_F$ is essentially 'round', and not elongated along $k_{y'}$ as along 
this direction one leaves the relevant mirror plane and thereby relaxes the symmetry selection rules that 
form part of the matrix elements. The same goes for the fact that we do not see a sign of the deeper 
lying Fe $3d_{x^2-y^2}$ states in either Fig. 11(b) or (d), although we do remark that the scattering rates at 
these energies are possibly already rather high and thus these bands cannot be resolved from the incoherent 
background, which may be substantial due to the Sn impurity scattering.

For the analogous data taken with horizontal (i.e. p) polarization, shown in Figs.~\ref{20KhoriM} and \ref{300KhoriM},
the odd Fe $3d_{x^2-y^2}$ states should disappear near the $k_{y'}$ = 0 line but the Fe $3d_{xz}$ and 
Fe $3d_{yz}$ states should remain visible. In both figures, the higher binding energy intensity 
distribution now no longer takes on the form of a blade-like structure parallel to $k_{x'}$, but 
rather one parallel to $k_{y'}$, as would be expected for the Fe $3d_{xz}$ and Fe $3d_{yz}$ states 
upon inspection of Fig. 5. For the data at $E_F$, the fact that the high intensity seems more 
concentrated close to $k_{x'}$=$k_{y'}$=0 with respect to the data of Fig. 11(a) would offer support for the DFT prediction 
of a narrower Fermi surface cross-section parallel to $k_{x'}$ for the Fe $3d_{xz}$ and Fe $3d_{yz}$ related states.
It remains a moot point whether the low temperature data for energies at or close to $E_F$ show signs 
of an elliptical intensity distribution with its long axis along $k_{y'}$ or not.
In any case, what is highly evident from both the low temperature (Fig. 9) and room temperature (Fig. 10) 
data from the X-point is that the intensity for positive $E-E_F$ values is much more robust, being clearly 
visible - even for the 20K data - up to energies some 80 meV above $E_F$. This is in stark contrast to the 
data of Fig. 6, taken around the $\Gamma$-point, whereby already only 40 meV above $E_F$ all intensity has 
disappeared. These facts serve as a clear indication of the electron-pocket nature of the states near X, and 
at the same time of the fact that the top of the band forming the hole pocket at $\Gamma$ is not far above 
the Fermi level in these samples.

It is interesting to note 
that even for temperatures well below
the transition temperature from the tetragonal paramagnetic state to 
the orthorhombic antiferromagnetic state, we see clear Fermi
surfaces in Figs.~\ref{20KhoriM} to \ref{20KvertM}, which would indicate that there is no massive gapping of 
the Fermi surface going on, due to a nesting between the Fermi surfaces at the
$\Gamma$ and the X-point. If such a gapping exists, we are not able to resolve it in the present data. These results are 
again in agreement with previous ARPES 
studies\cite{Yang2008} on BaFe$_2$As$_2$. On the other hand the present
results are at variance with
ARPES data\cite{Zabolotnyy2008}
measured on paramagnetic, K-doped BaFe$_2$As$_2$ in which a splitting of the bands 
at the X-point due to nesting was proposed. 

At the end of this Section we discuss the photon energy dependent data recorded around the $\Gamma$-point, 
shown in Fig. 12. The clearest conclusions from these data involve the clear presence or absence 
of spectral weight with a band-bottom at $E-E_F$ = -600meV. As discussed earlier in the light of 
our exploitation of the symmetry selection rules valid for the use of linear vertical and linear 
horizontal synchrotron radiation, these states have mainly Fe $3d_{z^2}$
character. All data in Fig. 12 are recorded with p-polarised light, so the mirror 
symmetry of these states means that they can be detected. For the panels of Fig. 12 taken with photon energies 
of 50, 75, 125 and to a slightly lesser extent 175 eV, these Fe $3d_{z^2}$ states are clearly seen with a 
binding energy of some 600 meV for $k_{y}$=0. This binding energy value is, in itself, a clear sign that, for 
these photon energies, we are probing states with $k_z$=0, i.e., we are probing the $\Gamma$-point or symmetry-equivalent 
points in higher Brillouin zones. For the photon energies of 100 and 150 eV, the Fe $3d_{z^2}$ states at 600meV 
are missing, indicating that their binding energy is very different (a $k_z$-dependence of the energy of a band 
built up of out-of-plane $3d_{z^2}$ states is not unreasonable) either/or there are strong matrix element differences 
sensitive to the $k_z$ value that prohibit their detection.  
If we take a value of 13 eV for the inner potential (a value derived from the bottom
of the valence band, taken from the data from the band structure calculations), we come to the $k_z$ values 
shown in Fig.~\ref{kz}. For an effective period along the z-direction corresponding to
the distance between two FeAs layers ($c'=c/2$), the panels for h$\nu$ = 50, 75, 125, and 175 eV would then 
correspond to $k_z$ values close to zero (or even multiples of $\pi/c'$), while the panels for h$\nu$ = 100 and 150 eV
would correspond to $k_z$ values close to $\pi/c'$ or odd multiples thereof. This would place the latter two 
data sets close to the Z-point (see Fig. 1).

In this manner, the analysis of the data of Fig. 12 underpins our earlier discussion, in which we 
compared the $k_z$=0 DFT band structure and constant energy surfaces (in Figs. 3, 4 and 5) with the ARPES data 
recorded with 75 eV photons. Furthermore, the $k_z$ assignment made possible by the data of Fig. 12, combined with the 
inspection of the data themselves near $E_F$ and  $k_{y}$=0 make it clear that the fairly 'closed' 
nature of the observed hole pocket near $\Gamma$ indicates that it comes from the innermost Fermi 
surface predicted by DFT. The fact that the data panels in Fig. 12 recorded with photon energies of 100 and 150 eV 
do not show a grossly different 'open' or 'closed' nature of this hole pocket is not inconsistent with the fact 
that the DFT calculations for the innermost surface predict only a 40\% variation in $k_F$ between $\Gamma$ and Z, 
compared to a 210\% increase in $k_F$ for the outermost $\Gamma$-centered hole pocket between $\Gamma$ and Z. 

Finally, despite some variation in the exact values of the photon energies used, we nevertheless 
note that the $k_z$ assignments for the h$\nu$ = 50 and 75 eV data as being close to $\Gamma$, 
match reasonably with other ARPES results regarding the locations of the $\Gamma$ and Z-points 
as a function of photon energy.\cite{Vilmercati2009}

\section{\label{sec:con}  CONCLUSIONS}

In this ARPES study we have determined the electronic structure of BaFe$_2$As$_2$, a parent compound
of doped FeAs-based high-$T_c$ superconductors. The hole (electron) pocket nature of the frontier electronic 
states at $\Gamma$(X) has been confirmed. Experimental information 
has been derived both for the paramagnetic tetragonal state and the antiferromagnetic 
orthorhombic state. No significant changes could be resolved between the two phases, indicating that the 
real differences for the electronic structure between these two phases are probably much smaller
than those predicted by spin-dependent DFT band structure calculations, a fact that supports a more itinerant 
character for the Fe 3d conduction electrons. 
In general, reasonable qualitative agreement between the band structure calculations 
for the paramagnetic state and the ARPES data has been observed, although in a number of points a direct 
assignment of the experimental photoemission intensity distributions to individual bands/Fermi surfaces is not without 
a residual level of ambiguity.
Photon energy dependent measurements have enabled a clear identification of which photon energies 
(for experiments around normal emission) give access to $k_z$ values at or near to the center of the 
Brillouin zone ($\Gamma$) and which probe closer to the zone boundary in $k_z$ (Z). 
The same data suggest modest three-dimensionality for the innermost Fermi surface (with respect to 
the $\Gamma$Z line), which is in keeping with the DFT predictions.
\section*{ACKNOWLEDGMENTS}
Financial support by the DFG is appreciated by J.F. (Forschergruppe FOR 538) and by R.V. and H.O.J. (SFB/TRR49, 
Emmy Noether program). This work is part of the research programme of the 'Stichting voor Fundamenteel Onderzoek 
der Materie (FOM)', which is financially supported by the 'Nederlandse Organisatie voor Wetenschappelijk Onderzoek (NWO)'. 

\bibliographystyle{apsrev}
\bibliography{BaFe2As2condmat}

\end{document}